\documentclass[runningheads]{llncs}
\usepackage{graphicx}
\usepackage{todonotes}
\usepackage{hyperref}
\begin{document}
\title{Participatory Design Landscape for the Human-Machine Collaboration, Interaction and Automation at the Frontiers of HCI (PDL 2021)\thanks{Supported by Human Aspects in Software Engineering (HASE) Research Initiative}}
%
%
\author{Wies\l{}aw Kope\'{c}\inst{1}\orcidID{0000-0001-9132-4171} \and
Cezary Biele\inst{2}\orcidID{0000-0003-4658-5510} \and
Monika Kornacka\inst{3}\orcidID{0000-0003-2737-9236} \and
Grzegorz Pochwatko\inst{4}\orcidID{0000-0001-8548-6916} \and
Anna Jaskulska\inst{5}\orcidID{0000-0002-2539-3934} \and
Kinga Skorupska\inst{1,3}\orcidID{0000-0002-9005-0348} \and
Julia Paluch\inst{1}\orcidID{0000-0002-7657-7856} \and
Piotr Gago\inst{1}\orcidID{0000-0001-7288-4210} \and
Barbara Karpowicz\inst{1}\orcidID{0000-0002-7478-7374} \and
Marcin Niewi{\'n}ski\inst{1}\orcidID{0000-0002-6416-3541} \and
Rafa\l{} Mas\l{}yk\inst{1}\orcidID{0000-0003-1180-2159}}

\authorrunning{Kopeć et al.}
\titlerunning{Participatory Design Landscape at the Frontiers of HCI}
%
\institute{Polish-Japanese Academy of Information Technology \and National Information Processing Institute \and
SWPS University of Social Sciences and Humanities\\
\and IP Polish Academy of Sciences \\
\and KOBO Association}
\maketitle              
\begin{abstract}
We propose a one-day transdisciplinary creative workshop in the broad area of HCI focused on multiple opportunities of incorporating participatory design into research and industry practice. This workshop will become a venue to share experiences and novel ideas in this area. At the same time, we will brainstorm and explore frontiers of HCI related to engaging end users in design and development practices of established and emerging ICT solutions often overlooked in terms of co-design. We welcome a wide scope of contributions in HCI which explore sustainable opportunities for participatory design and development practices in the context of interconnected business, social, economic and environmental issues. The contributions ought to explore challenges and opportunities related to co-design at the frontiers of HCI - participatory design of newest and complex technologies, not easily explainable or intuitive, novel collaborative (remote or distributed) approaches to empowering users to prepare them to contribute as well as to engaging them directly in co-design.

\end{abstract}

\section{Theme and Topics}

\subsection{Theme}
Our workshop on participatory design landscape for the human-machine collaboration, interaction and automation welcomes contributions in the broad area of Human-Computer Interaction. We focus on facilitating end-users involvement in a sustainable and ethical way in the context of the novel ICT solutions, remote co-design participation and accelerating digital transformation touching both the professional and the private spheres. This transformation presents multiple opportunities for addressing complex interdependent problems with co-designed ICT-based solutions using remote or distributed collaboration and novel empowerment approaches, to enable all users to fully participate in co-designing even complex solutions, which are not easy to relate to or intuitive.


\begin{figure}[!ht]
  \centering
  \includegraphics[width=\linewidth]{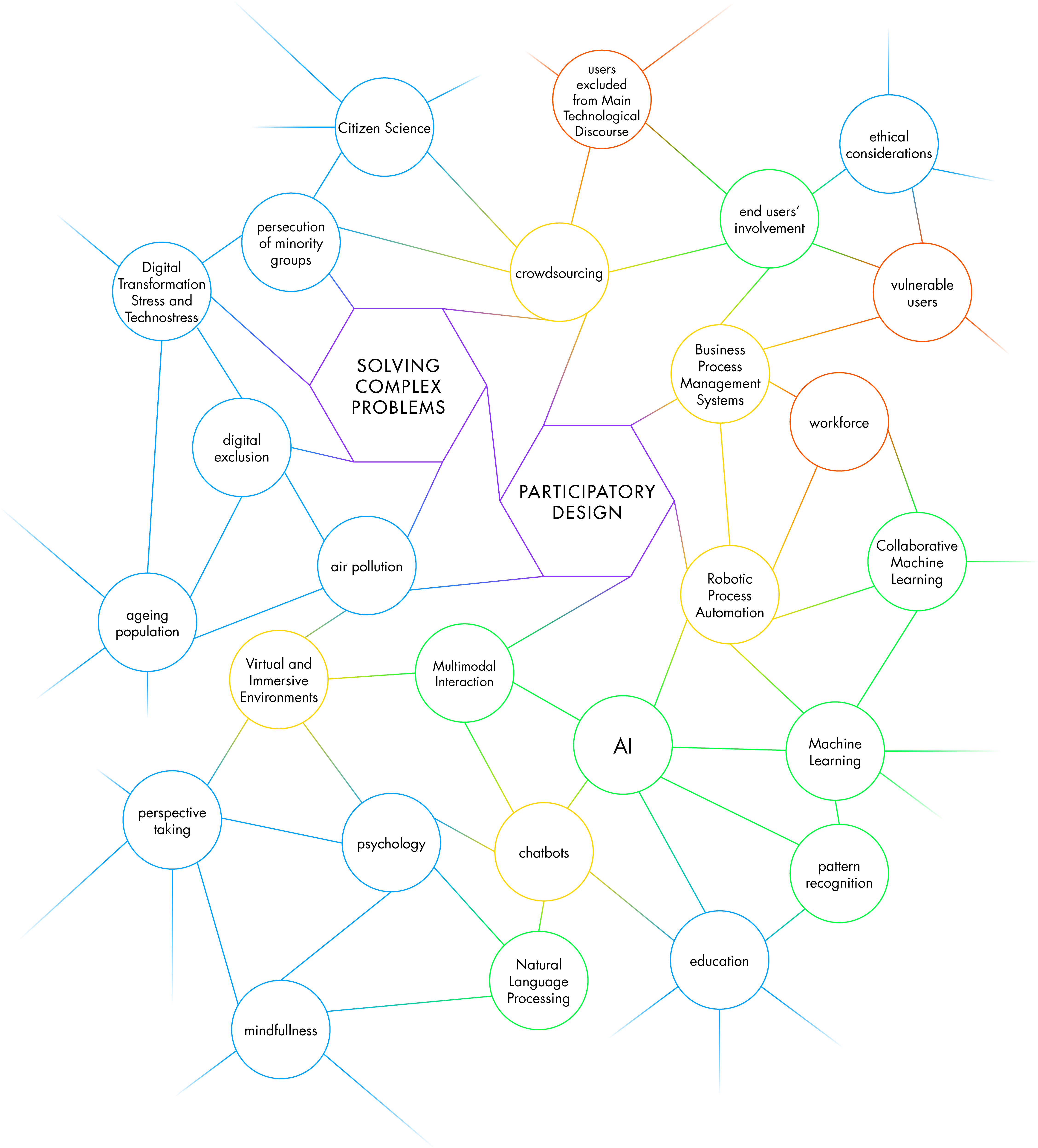}
  \caption{Example topics for this workshop in relation to its key co-design theme.}
  \label{mindmap}
\end{figure}

\subsection{Contributions}

The contributions ought to explore challenges and opportunities related to co-design at the frontiers of HCI - participatory design of newest and complex technologies, not easily explainable or intuitive, novel collaborative (remote or distributed) approaches to empowering users to prepare them to contribute as well as to engaging them directly in co-design. Therefore, we welcome reports of interdisciplinary projects, studies and potential research initiatives which use (or would like to benefit from) various participatory approaches. These can, for example, include cases of user empowerment \cite{davidson2013participatory,sanders2002user,kopec2017spiral,ladner2015design} or engaging users in diverse co-design contexts, such as Living Labs \cite{pallot2010living,kopec2017living}, design probes, and other creative co-design methods. 


\subsection{Topics}
These studies, while necessarily focusing on participatory design practices, can touch upon various aspects of applied computer science (including data science, artificial intelligence, social informatics and human-computer interaction) alongside with current state-of-the-art in psychology, social science and economics (such as neuroscience, psychophysiology and cognitive studies). The interdisciplinary and participatory approach they ought to describe can be applied in the context of interconnected business, social, economic and environmental issues, as our focus is on the use of co-design approaches in a wide scope of novel contexts. Therefore, in the workshop we will explore the opportunities related to participatory design of both established and emerging solutions and interfaces like Voice Assistants (VA), Virtual (VR), Augmented (AR) and Mixed Reality alongside with major underlying trends like Artificial Intelligence (AI), Machine Learning (ML), Natural Language Processing (NLP) or Complex Event Processing (CEP) in the context of multiple areas of interest to science and industry where co-design can be particularly beneficial. One example of such area of interest is related to creating health solutions using technologies that are new to end users - patients - but that have already been proven to be effective at improving the efficiency of mental health therapy, such as VR or AR. Another opportunity is related to the co-development of self-help mobile applications or games \cite{GamefulmHealth2018} for health (including mental health) \cite{MyStrengths2020} and of wearable monitoring solutions. The development and increased access to wearable, mobile and robotic technology brings novel opportunities for using IT-based solutions in everyday life, including personal spaces and workplaces. Such solutions make it possible to measure psychological processes and to provide treatment in ecological conditions of patients' everyday life. Yet, research and practice show that functionality and ergonomics of the proposed solutions are crucial for acceptability and compliance. Thus, participatory design may be an adequate way to address those challenges. Especially the design of social robots, which heavily rely on expectations based on social interaction norms, stands to greatly benefit from participatory design \cite{piccarra2016designing}.

\section{Organizers}

We are a multidisciplinary group of scientists, researchers, IT professionals and activists whose key focus is on multiple positive applications of ICT through engaging in participatory practices\footnote{\href{https://www.youtube.com/watch?v=mRaS1Gn_ZvE&t=1s}{Participatory Design Workshops}}, especially focused on engaging the users excluded from the technological discourse \cite{kopec2017spiral} (for example, with games\footnote{\href{https://www.youtube.com/watch?v=AdF-YMgOlSo&feature=youtu.be}{Location-Based Game}} or hackathons\footnote{\href{https://www.youtube.com/watch?v=v3lFz3z_5Hw&feature=youtu.be}{Hackathon}} or through inviting stakeholders to take part in educational and capacity building projects\footnote{\href{https://alienproject.pja.edu.pl/}{ALIEN Project}}).

\subsubsection{Key Organizers}
\begin{itemize}
\item Wies\l{}aw Kope\'{c}, PhD, MBA, 
(\href{https://scholar.google.fi/citations?user=aFYddKYAAAAJ&hl=en}{Google Scholar Profile})
\\Head of XR Lab, Computer Science Department, 
\\Polish-Japanese Academy of Information Technology (PJAIT) \\
\item Cezary Biele, PhD, (\href{https://scholar.google.fi/citations?user=3DNnWwQAAAAJ&hl=en}{Google Scholar Profile})
\\Head of Laboratory of Interactive Technologies, 
\\National Information Processing Institute (NIPI)\\
\item Monika Kornacka, PhD, (\href{https://scholar.google.fi/citations?user=RzJuLgEAAAAJ&hl=en}{Google Scholar Profile})\\
Head of Emotion Cognition Lab, Institute of Psychlogy, 
\\SWPS University of Social Sciences and Humanities (SWPS)\\
\item Grzegorz Pochwatko, PhD, (\href{https://scholar.google.fi/citations?user=OzQwZ04AAAAJ&hl=en}{Google Scholar Profile})
\\Head of Virtual Reality and Psychophysiology Lab, Institute of Psychology, Polish Academy of Sciences (IP PAS)
\end{itemize}

\subsubsection{Program Committee}
Jakub Mo\.{z}aryn, PhD, Warsaw University of Technology (\href{https://scholar.google.fi/citations?user=TxNaH70AAAAJ&hl=en}{Google Scholar Profile});
Jacek Lebied\'{z}, PhD, Gdansk University of Technology (\href{https://pg.edu.pl/4fc28b7093_jacek.lebiedz}{University Profile});
Marzena Wojciechowska, PhD, PJAIT Resarch Center (\href{https://www.researchgate.net/profile/Marzena_Wojciechowska2}{ResearchGate Profile});
Dominika Tkaczyk, PhD, Crossref (\href{https://scholar.google.fi/citations?user=pk9aObIAAAAJ&hl=en}{Google Scholar Profile});
Sebastian Zagrodzki, Google (\href{https://ch.linkedin.com/in/sebastianzagrodzki}{LinkedIn Profile})
\L{}ukasz Czarnecki, Amazon (\href{https://pl.linkedin.com/in/lukasz-czarnecki-95262a39}{LinkedIn Profile});
Anna Jaskulska, Kobo Association and Living Lab (\href{http://kobo.org.pl/}{Kobo Profile});
Ewa Makowska, Business and System Analyst and IT Project Manager (\href{https://pl.linkedin.com/in/ewa-makowska-tlomak-3801134}{LinkedIn Profile}).

\subsubsection{Paper Co-Chairs}
Kinga Skorupska, XR Lab PJAIT, (\href{https://scholar.google.fi/citations?user=zEugAPUAAAAJ&hl=en}{Google Scholar Profile}) and
Julia Paluch, XR Lab PJAIT (\href{https://www.researchgate.net/profile/Julia-Paluch}{ResearchGate Profile}).
\subsubsection{Technical Chairs}
Rafa\l{} Mas\l{}yk, XR Lab PJAIT; Barbara Karpowicz, XR Lab PJAIT; Maciej Krzywicki, XR Lab PJAIT; Piotr Gago, XR Lab PJAIT.

\section{Objectives}
In organizing this workshop we have three key objectives:
\begin{enumerate}
\item Exchanging experiences, ideas, best practices and guidelines for co-designing at the transdisciplinary frontiers of HCI.
\item Outlining the participants' aspirations and opportunities for breaking new ground in the area of HCI and exploring ideas for innovative project proposals and practices involving participatory design.
\item Establishing common ground and voice for the discussion of the challenges and opportunities related to the future of participatory practices at the frontiers of HCI.
\end{enumerate}

\section{Target Audience}
We invite researchers and practitioners in the broad area of HCI, who want to explore the opportunities of incorporating participatory practices in their work, including design, collaborative ML, RPA, and crowdsourcing in an ethical and sustainable way. At the same time we want to find new areas where co-design of ICT-based solutions may work towards improving social cohesion and addressing complex business, social, economic and environmental issues.

\section{Methodology}

We propose a one-day exploratory workshop combining sharing best practices as interactive presentations and discussions with a creative flair of a collaborative design-inspired toolbox geared towards exploring opportunities at the frontiers of co-design and HCI. In the course of the workshop we want to use creative tools to aid free thinking, instead of stifling it with over-formalized "creative" methodology focused on structured results. We believe that allowing for a degree of ambiguity at the beginning of the creative process is a must, while the outlines of our combined expertise are sufficient to work as bounds that limit the scope of our imagination. In this vein, the workshop will be led in a friendly atmosphere, without formalized introductions or strict presentation formats. Our workshop facilitators have diverse creative tools and practices at their disposal, which they can adjust accordingly to the number and needs of the participants and their levels of energy. 

\section{Expected Outcomes}

As our key motivation is to explore the opportunities related to participatory design at the frontiers of HCI, while grounding these firmly in established and emerging practices, the most important outcome will be a post-workshop multi-author publication. Its format will be the result of issues discussed during the workshop, but the idea for it is inspired by Kittur et al. \cite{kittur2013future} whose paper "The Future of Crowd Work" clearly outlined the aspects, trends and opportunities related to crowdsourcing. We feel the same is needed in the area of participatory design and human factors at the frontiers of HCI, as such design ought to take into account ethics, preferences and motivations as well as the broad user experience and visual design. Especially now, that more remote tools and approaches are needed to engage users, and novel forms of empowerment ought to emerge to let co-designers really understand the context of the complex co-designed solutions. Additionally, we would be happy to invite our participants to look for opportunities to publish a collection of extended papers from the workshop.

\bibliographystyle{splncs04}
\bibliography{mybibliography}

\end{document}